\documentclass[preprint,secnumarabic,tightenlines,amssymb,amsmath,nobibnotes,aps,nofootinbib,showpacs]{article}%
\usepackage{graphicx}
\usepackage[caption=false]{subfig}
\usepackage{dcolumn}
\usepackage{bm}
\usepackage{amssymb}
\usepackage{amsmath}
\usepackage{amsfonts}
\usepackage{verbatim}
\usepackage{hyperref}
\usepackage{color}
\usepackage[parfill]{parskip}

\providecommand{\U}[1]{\protect\rule{.1in}{.1in}}
\definecolor{darkgreen}{rgb}{0,0.35,0}

\definecolor{Rood}{rgb}{1, 0, 0}
\date{}
\begin{document}
\title{\noindent\textbf{The Gribov problem in presence of background field for $SU(2)$ Yang-Mills theory}}
\author{Fabrizio Canfora${}^{1}$\thanks{canfora@cecs.cl}, Diego Hidalgo${}^{1,2}$\thanks{dhidalgo@cecs.cl}, Pablo Pais${}^{1,3}$\thanks{pais@cecs.cl}\\$^{1}$\textit{Centro de Estudios Cient\'{\i}ficos (CECS), Casilla 1469,
Valdivia, Chile.}\\$^{2}$\textit{ Departamento de F\'{\i}sica, Universidad de Concepci\'on,
Casilla 160, Concepci\'on, Chile}\\$^{3}$\textit{ Physique Th\'eorique et Math\'ematique,Univ\'ersite de
Bruxelles and }\\\textit{International Solvay Institutes, Campus Plaine C.P. 231, B-1050
Bruxelles, Belgium}}
\maketitle

\begin{abstract}
The Gribov problem in the presence of a background field is analyzed: in
particular, we study the Gribov copies equation in the Landau-De Witt gauge
as well as the semi-classical Gribov gap equation. As background field, we
choose the simplest non-trivial one which corresponds to a constant gauge
potential with non-vanishing component along the Euclidean time direction.
This kind of constant non-Abelian background fields is very relevant in
relation with (the computation of) the Polyakov loop but it also appears when
one considers the non-Abelian Schwinger effect. We show that the Gribov copies
equation is affected directly by the presence of the background field, constructing an explicit example. The analysis of the Gribov gap
equation shows that the larger the background field, the smaller the Gribov mass parameter. These results strongly suggest that the relevance of the Gribov copies (from the path integral point of view) decreases as the size of the background field increases.
\end{abstract}
\maketitle

\section{Introduction}

The main tool to compute observable quantities in QFT is perturbation theory.
In gauge theories, and in Yang-Mills (YM) theory in particular, a fundamental
problem to solve in order to compute physical quantities is the over-counting
of degrees of freedom related to gauge invariance (for a detailed analysis see
\cite{faddeev-popov}). The Faddeev-Popov (FP) gauge fixing procedure is the
cornerstone which allows using the Feynman rules and Feynman diagrams in all
applications of the standard model. The obvious fundamental hypothesis is that
the gauge-fixing condition must intersect once and only once every gauge
orbit. Locally, in the space of gauge fields, this hypothesis requires that the
FP operator should not have zero modes so that the FP
determinant is different from zero. The reason is that the existence of a
proper gauge transformation preserving the gauge-fixing would spoil the whole
quantization procedure since it would imply that the FP recipe does not
completely eliminate the over-counting of degrees of freedom.

However, in \cite{gribov}, Gribov showed that in non-Abelian gauge theories
(in flat, topologically trivial space-times) the FP procedure fails
at non-perturbative level. The reason is that a proper gauge fixing is not
possible due to the appearance of Gribov copies: namely, gauge equivalent
configurations satisfying the Coulomb gauge. Later, Singer \cite{7} showed
that if Gribov ambiguities occur for the Coulomb gauge, they occur for all
gauge fixing conditions involving derivatives of the gauge field.

Naively, one could expect to completely avoid the Gribov problem by simply
choosing algebraic gauge fixings like the axial gauge or the temporal gauge,
which are free of Gribov copies. However, these choices have their own, and even
worse, problems\footnote{The origin of these problems is that in all these
algebraic gauges the free propagator of the gauge field is more singular than
in linear covariant ones owing to the presence of additional \textquotedblleft
spurious\textquotedblright\ singularities \cite{9}.} (for a detailed reviews
see \cite{8} \cite{11}). Here it is just worth mentioning one serious issue:
any loop computation in the algebraic gauge-fixings mentioned above are very
difficult already beyond two-loop. Hence, from the practical point of view,
linear covariant gauge-fixings are far more convenient: here we will only
consider this kind of gauge-fixing.

On the other hand, the existence of Gribov copies is not just a problem since,
as Gribov himself argued, the natural way to solve such a problem is able to
shed considerable light on the infrared (IR) region of YM
theory. Such solution is to restrict the path-integral only to a region
$\Omega$, which is called Gribov
region, where FP operator is definite-positive \cite{gribov,14,15,16,17} (detailed reviews are
\cite{notasdesorella} and \cite{11}) so that there are no Gribov copies connected to the
identity\footnote{Some Gribov copies are still left within the Gribov region
\cite{14}. One can define a \textit{modular region} which is completely free
of Gribov copies (both small and large). However, how to implement the
restriction to the modular region is not known yet. Thus, we will work within
the Gribov region as it is usually done.}. In order to restrict the path
integral to the Gribov region, one can use the
Gribov-Zwanziger (GZ) approach \cite{12,13}.
When the space-time geometry is flat and
the topology trivial\footnote{On the other hand, on curved spaces the pattern
of appearance of Gribov copies can be considerably more complicated (for
instance, even Abelian gauge theories can have `induced' gauge copies
\cite{nostro1,nostro2,nostro3,nostro4,decesare,nostro5,nostro5.5,CGZ}). Thus, in
the following only the standard flat case will be considered.}, this method is
able to reproduce the usual perturbation theory encoding, at the same time,
the effects related to the elimination of the Gribov copies. For instance, it
allows the computation of the glueball masses in excellent agreement with the
lattice data \cite{sorellaPRL,cucchieri,fischer}. Within the same framework,
it is also possible to solve the sign problem for the Casimir energy and force
in the MIT-bag model \cite{nostro6}.

This scheme works very well also at finite-temperature \cite{zwat1,
nostro7, zwat2,annals,fukushima} (this is also supported by the results in
\cite{loewepaper1}). Moreover, at one-loop order, it is possible to compute
the vacuum expectation value for the Polyakov loop \cite{polyakov-paper}:
these results are in a good agreement with the expected behavior for the
deconfinement phase transition \cite{fingber-heller-karsch}. Within the GZ
approach, the non-perturbative correction to the gluon propagator is encoded
in the Gribov mass which is determined in a self-consistent way by solving the
so-called Gribov gap equation. Therefore, the analysis of the dependence of
Gribov mass on the temperature (as well as on other relevant external
parameters) is very useful to determine the phase-diagram of Yang-Mills theory.

Thus, it is natural to wonder whether or not this approach works so well also
in the presence of a background gauge field. From the theoretical point of
view, this analysis is very important as it discloses how strongly the
presence of a background field can affect the Gribov region and the whole
issue of Gribov copies. One of the most relevant applications of the
background field method is the computation of the (vacuum expectation value of
the) Polyakov loop \cite{polyakov-paper} in which the presence of the Polyakov
loop manifests itself as a constant background field with component along the
Euclidean time\footnote{Formally, a constant background gauge field with only
the timelike component non-vanishing is related to a bosonic chemical
potential \cite{kapustamethod,paper-argument-thermodynamic}. On the
other hand, the physical interpretation of such Bosonic chemical potential is
rather obscure in the case of non-perturbative gluons and so it will not be
discussed in the present case.}. Another very important non-perturbative
phenomenon in which the presence of a background gauge field plays a key role
is the (both Abelian and non-Abelian) Schwinger effect \cite{schwinger1,schwinger2,schwinger3}. Also in the case of the non-Abelian
Schwinger effect, the relevant background
gauge fields are constant $A_{\mu}$ which have components both along time and
space directions. From the point of view of applications, such an analysis can
also be quite relevant in relation with quark-gluon plasma
\cite{QG-plasma1,QG-plasma2}, color superconductivity in QCD
\cite{superconductivity-QCD}, astrophysics \cite{astrophys1,QG-plasma2}, and
cosmology \cite{cosmology1,cosmology2}.

The idea of the present paper is precisely to begin the study of the following
very relevant and broad question: how the presence of a background field
affects the (appearance of) Gribov copies as well as the gap equation form. To the
best of authors knowledge, such issue has not been deeply analyzed so far.

The Background Field Method (BFM) \cite{dewitt67,thooft,abbott82} together with the techniques developed in
\cite{nostro1,nostro2,nostro3,nostro4,decesare,nostro5,nostro5.5} are adopted
in the present paper. The results of these references on the Gribov problem on
curved space strongly suggest (taking into account that the background metric
can play the role of an external field) that background fields can play a
prominent role within the GZ approach to YM theory. Here we show that the Gribov copies equation is affected directly by the
presence of a background field. In particular, explicit examples will be
constructed in which the ``relevance" of the allowed Gribov copies decreases as
the background field is increased. Moreover, the analysis of the
semi-classical Gribov gap equation shows that the Gribov mass parameter
decreases as the size of the background field is increased.\newline

The paper is organized as follows. In the second section, the Gribov problem
in the Landau-De Witt gauge is introduced. In the third section, explicit
examples of Gribov copies in the Landau-De Witt gauge are studied. In the
fourth section, the Gribov gap equation within a background field is analyzed.
Some conclusions and discussions are drawn at the end.

\section{A brief review of the Gribov-Zwanziger action}\label{GZ-action-section}

In this section we present an outlook of the GZ-approach without background consider background fields, which is the aim of Section \ref{GZ-background}. The Euclidean Yang-Mills action
\begin{equation}\label{YM-action}
S_{YM} = \frac{1}{4} \int d^4x F^{a}_{\mu \nu} F^{a \mu \nu}, \quad F_{\mu\nu}^{a}=\partial_{\mu}A_{\nu}^{a}-\partial_{\nu
}A_{\mu}^{a}+gf^{abc}A_{\mu}^{b}A_{\nu}^{c},
\end{equation}
is invariant under the gauge transformation
\begin{equation}\label{symmmetry-YM}
A_\mu \rightarrow A_\mu = U^{-1}A_\mu U + U^{-1}\partial_\mu U,
\end{equation}
with $U \in SU(N)$. In order to take into account the existence of Gribov copies due to the this gauge transformation, Gribov proposed \cite{gribov} to restrict the domain of integration in the path integral to a region in
functional space where the eigenvalues of the FP operator $\mathcal{M}^{ab}$
are strictly positive. This region is known as the Gribov region $\Omega$, and
is defined as
\begin{equation}\label{Gribov-region}
\Omega=\{A_{\mu}^{a}\; | \; \quad\partial_{\mu}A_{\mu}^{a}=0;\quad\mathcal{M}%
^{ab}=-\partial_\mu (\partial^{\mu}\delta^{ab}-gf^{abc}A_{\mu}^{c})=-\partial_\mu D^{ab}_{\mu}>0\}.
\end{equation}
where $ D^{ab}_{\mu} = \partial^{\mu}\delta^{ab}-gf^{abc}A_{\mu}^{c}$ is the usual covariant derivative, which depends on $A_{\mu}$. The boundary of this region is called the first Gribov horizon. Later on, Zwanziger \cite{12} implemented the Gribov region $\Omega$ in Euclidean Yang-Mills theories in the Landau gauge, by means of the following action
\begin{equation}\label{Yang-Mills-action}
S_h = S_{YM} + \int d^4x \left( b^a \partial_\mu A^{a}_{\mu} + \bar{c}^a \partial_\mu D^{ab}_{\mu}c^b   \right) + \gamma^4 \int d^4 x h(x),
\end{equation}
with $S_{YM}$ the Euclidean version of the Yang-Mills action defined in \eqref{YM-action}, and where $h(x)$ is the so-called horizon function
\begin{equation}
h(x) = g^2 f^{abc} A^{b}_{\mu} (\mathcal{M}^{-1} )^{ad} f^{dec} A^{e}_{\mu}.
\end{equation}
The $\gamma$ parameter, known as Gribov mass parameter, at a semi-classical
level, provides a detailed description of the confinement as the poles of the
propagators are imaginary when $\gamma^{2}\neq0$ \cite{notasdesorella}, and is
determined by a self-consistent horizon condition
\begin{equation}\label{HC1}
\langle h(x) \rangle  = d(N^2 -1 ),
\end{equation}
where $d$ is the number of the space-time dimensions and we understand for $\langle\ldots\rangle$ functional integral over the fields. The local version of the horizon function $h(x)$ can be achieved through a suitable set of additional fields, which belong to a BRST doublet. Then, the local full action reads,
\begin{align}
S_{GZ}  &  =S_{FP}+S_{\gamma}+S_{0},\label{GZaction}\\
S_{FP}  &  =\int d^{4}x\left(  \frac{1}{4}F_{\rho\sigma}^{a}F^{\rho\sigma
a}+ib^{a}\partial_{\rho}A_{\rho}^{a}+\bar{c}^{a}\mathcal{M}^{ad}c^{d}\right)
,\label{FP-action}\\
S_{\gamma}  &  =\int d^{4}x\left(  \gamma^{2}gf^{abc}A_{\rho}^{a}%
(\varphi_{\rho}^{bc}-\bar{\varphi}_{\rho}^{bc}) \right) + \gamma^4 \int d^4 x h(x)
,\label{gamma-action}\\
S_{0}  &  =\int d^{4}x\left(  -\bar{\varphi}_{\rho}^{ac}\mathcal{M}%
^{ab}\varphi_{\rho}^{bc}+\bar{\omega}_{\rho}^{ac}\mathcal{M}^{ab}\omega_{\rho
}^{bc}+gf^{amb}(\partial_{\rho}\bar{\omega}_{\sigma}^{ac})(D_{\rho}^{mp}%
c^{p})\varphi_{\sigma}^{bc}\right)  ,
\end{align}
where Greek indexes run from $\mu=1\ldots d$ and latin indexes from $a=1\ldots N^{2}-1$.
The fields $(\bar{\varphi}^{ac}_{\mu} , \varphi^{ac}_{\mu})$ are a pair of complex conjugate bosonic fields, while $(\bar{\omega}^{ac}_{\mu}, \omega^{ac}_{\mu})$ are anti-commuting fields.
\par
Now, if we consider the relation between the local action $S_{GZ}$ and the non-local action $S_h$
\begin{equation}
\int [dA] [db] [dc] [d \bar{c}]e^{-S_h} = \int [dA] [db] [dc] [d \bar{c}] [d \varphi ]d[ \bar{\varphi}]
d \omega] [d \bar{\omega}]    e^{-S_{GZ}}
\end{equation}
and we take the partial derivative of both sides w.r.t $\gamma^2$ (with $\gamma \neq$ 0), we obtain
\begin{equation}
\langle g f^{abc}A^{a}_{\mu} \varphi^{bc}_{\mu} \rangle + \langle g f^{abc}A^{a}_{\mu} \bar{\varphi}^{bc}_{\mu} \rangle + 2 \gamma^2 d (N^2 -1) = 0 ,
\end{equation}
which it is precisely the horizon condition \eqref{HC1}. On the other hand, we know that the effective action $\varepsilon_{vac}$ is obtained through
\begin{equation}
e^{-\varepsilon_{vac}} = \int \left[d \Phi\right] e^{-S_{GZ} },
\end{equation}
where $\int [d \Phi]$ stands for the integrations over all the fields contained in the action $S_{GZ}$. From this last expression, the $\gamma$ parameter can also be determined by a self-consistent way by the following gap equation
\begin{equation}
\frac{\partial\varepsilon_{vac}}{\partial\gamma^{2}}=0.
\label{definition:gapequation}%
\end{equation}
Therefore, equation \eqref{definition:gapequation} represents the horizon condition formula which will allow us to determine the Gribov parameter later on.

\section{Gribov-Zwanziger action in a background field}\label{GZ-background}

As we present a brief introduction in Section \ref{GZ-action-section} of the GZ-approach, now we analyze what happens if we take into account a background field. We consider the $SU(N)$ Yang-Mills theory in $d=4$ Euclidean dimensions defined in Eq.\eqref{Yang-Mills-action}. In the BFM (see \cite{8,Weinberg-book} for more details), one introduces a fixed background gauge
field configuration $B_{\mu}$ through the splitting
\begin{equation}
A_{\mu}\rightarrow a_{\mu}\equiv A_{\mu}+B_{\mu}\ ,\label{BFreplacement}%
\end{equation}
where $A_{\mu}$ and $B_{\mu}$ play completely different roles. On the one hand, $A_{\mu}$
represents the quantum fluctuations of the gauge field. On the other hand, the
background field $B_{\mu}$ plays the role of a classical background, (this approach
is quite relevant in the case of the Polyakov loop computation
\cite{polyakov-paper}). The gauge symmetry \eqref{symmmetry-YM} changes with this background field as
\begin{equation}
A_{\mu}+B_{\mu}\rightarrow A_{\mu}^{\prime}+B_{\mu}^{\prime}=U^{-1}%
\partial_{\mu}U+U^{-1}\left(  A_{\mu}+B_{\mu}\right)
U\ .\label{backgroundsymmetry}%
\end{equation}
Although it is not mandatory, in many applications (such as the already
mentioned case of the Polyakov loop) it is convenient to demand that the
background gauge field is fixed (namely, it does not transform under gauge
transformations \eqref{symmmetry-YM}). Consequently, it follows the natural requirement
\[
\delta B_{\mu}^{a}=0\ .
\]
\ In this case, the symmetry transformation in Eq. (\ref{backgroundsymmetry})
can be written as%
\begin{equation}
A_{\mu}\rightarrow A_{\mu}^{U}=U^{-1}\partial_{\mu}U+U^{-1}A_{\mu}U+\left(
U^{-1}B_{\mu}U-B_{\mu}\right)  \ ,\label{backgroundsymmetry1}%
\end{equation}
where it has been explicitly taken into account that $B_{\mu}$ is not affected
by the gauge transformation. At the infinitesimal level, $U\approx
\mathbb{I}+\omega^{a}\tau_{a}$, $\omega\ll1$, one recovers the usual
infinitesimal gauge transformations with a background gauge field \cite{8,Weinberg-book}
\begin{align}
\delta A_{\mu}^{a} &  =f^{abc}\omega^{b}(A_{\mu}^{c}+B_{\mu}^{c})+\frac{1}%
{g}\partial_{\mu}\omega^{a},\label{infinitesimal-transf}\\
\delta B_{\mu}^{a} &  =0.
\end{align}
Correspondingly, the Landau gauge-fixing condition is also modified. In the presence of a background field, the most
convenient gauge-fixing condition takes the form
\begin{equation}
\tilde{G}_{\mu}^{a}[B]\equiv\overline{D}^{ab}_{\mu}A_{\mu}^{b}=0,\ \ \overline{D}_{\mu
}^{ab}:=\partial_{\mu}\delta^{ab}+gf^{acb}B_{\rho}^{c}\ ,
\label{backgroundgaugefixingcondition}
\end{equation}
known as the Landau-DeWitt (LDW) gauge fixing condition.
The FP procedure in the presence of a background gauge field leads to the
following action (see \cite{binosi-quadri} for a detailed discussion)
\begin{equation}
S_{B}^{gf}=\int d^{4}x\left(  \frac{1}{4}F_{\mu\sigma}^{a}F^{\mu\sigma
a}+\bar{c}^{a}\overline{D}_{\mu}(B)D_{\mu}(a)c^{a}-\frac{(\overline{D}_{\mu}(B)A_{\mu})^{2}%
}{2\xi}\right)
\end{equation}
with $c$ and $\bar{c}$ denoting the ghost and antighost fields, respectively.
The LDW gauge is actually recovered in the limit $\xi\rightarrow0$, taken at
the very end of each computation, and is also plagued by Gribov copies, as we will show in the following sections.
On the other hand, the GZ method can be applied to this situation by means a suitable choose of the background field $B_\mu$ in order to the new FP operator $\mathcal{\overline{M}}^{ac} \equiv - \overline{D}^{ab}_{\mu} (B)D^{bc}_{\mu}(a)$ is invertible inside the Gribov region $\Omega$. Following the lines of \cite{vandersickel-zwanziger} (and  \cite{polyakov-paper} in the case of a fixed background), the GZ action under the LDW gauge acquires the form
\begin{align}
S_{GZ} =\int& d^{4}x\left(  \frac{1}{4}F_{\mu\nu}^{a}F^{\mu\nu
a}+\overline{c}^{a}\overline{D}_{\mu}(B)D_{\mu}(a)c^{a}-\frac{\left(\overline{D}_{\mu}(B)A_{\mu}\right)^{2}%
}{2\xi}+\bar{\varphi}_{\mu}^{ac}\overline{D}_{\sigma}^{ab}(B)D_{\sigma}%
^{bd}(a)\varphi_{\mu}^{dc}\right.  \nonumber\label{backgroundactionGZ}\\
-&  \left.  \bar{\omega}_{\mu}^{ac}\overline{D}_{\nu}^{ab}(B)D_{\nu}%
^{bd}(a)\omega_{\mu}^{dc}-g\gamma^{2}f^{abc}A_{\rho}^{a}(\varphi_{\mu}%
^{bc}+\bar{\varphi}_{\mu}^{bc})-\gamma^{4}d(N^{2}-1)\right)  .
\end{align}
The equation \eqref{backgroundactionGZ} deserves some remarks. There are two key mathematical requirements (described in the references
\cite{gribov, 12}) necessary in order to write down a local GZ action which, in the presence of a background field, can be different with respect to the usual cases. The first one corresponds to the condition that the FP operator at tree-level must be invertible (otherwise, the whole procedure to
localize the horizon condition would be impossible as non-invertible operators would be involved). In the Landau gauge this is obvious as the FP operator at tree-level does not depend at all on $A_{\mu}$ and so it reduces to the
flat Laplacian (on curved spaces, the story can be quite different
\cite{nostro1,nostro2,nostro3,nostro4,decesare,nostro5,nostro5.5}, but we will consider flat spaces only in this
manuscript). In the cases in which there is a background field and one adopts
the LDW gauge-fixing condition, as we do here, such requirement becomes
$\overline{D}(B)_{\mu}D(a)^{\mu}$ must be invertible.
In other words, the above condition is just the statement that the vacuum belongs to the Gribov region, Such requirement is quite non-trivial in presence of background fields and it is possible to construct examples in which it is not satisfied. However, in the cases of the background fields considered in this paper (which are relevant in relation with both the Polyakov loop and the Schwinger effect computations) the above condition is satisfied.

The second key requirement (which does change in the presence of a background
field) is the validity of the \textit{Dell`Antonio-Zwanziger} theorem \cite{17}. In the case of the Landau gauge, such a theorem
provides the whole Gribov-Zwanziger idea with solid bases since it shows that
one does not loose any relevant information when the restriction to the Gribov
region is implemented (since \textit{Every Gauge Orbit Passes Inside the
Gribov Horizon}).\footnote{In the seminal paper, Gribov already had
the intuition that this powerful theorem might hold and he was able to prove
it locally (namely, for configurations which are very close to the horizon \cite{gribov}. A
nice review of the Gribov original argument is in section \textbf{3.3} of
\cite{notasdesorella}).} Remarkably, Gribov based its idea of the restriction to
the Gribov region on this \textit{local} version of the
\textit{Dell`Antonio-Zwanziger} theorem. In the presence of a background gauge
field, many of the technical assumptions of \cite{17} do not
hold in general. Consequently, the generalization of the
\textit{Dell`Antonio-Zwanziger} theorem appears to be a very difficult problem
in non-linear functional analysis (on which we hope to come back in a future
publication). On the other hand, the local argument by Gribov can be repeated
step by step in the case of the LDW gauge provided the background field is constant and commutes with itself (as is the case for the background field considered in the next section).

\section{The simplest non-trivial background field}
\label{section_gribov_background}

In order to describe the effects of a background field both on the Gribov
copies equation and on the gap equation avoiding unnecessary technical
complications, we will consider the simplest non-trivial background gauge
field (which is relevant in the computation of the Polyakov loop
\cite{polyakov-paper}):
\begin{equation}
B_{\rho}^{a}=\frac{r_{0}}{g}\delta^{a3}\delta_{\rho0}%
\ .\label{backgroundasMU}%
\end{equation}
\newline
Constant background non-Abelian gauge fields are very relevant in the analysis
of the non-Abelian Schwinger effect \cite{schwinger2,schwinger3} too.
However, the most interesting configurations considered in these works have both time-like and space-like components turned on at
the same time. Here we have chosen the above background gauge field with only
Euclidean time component since it allows to construct explicitly analytic
examples of Gribov copies as well as to solve the
semi-classical Gribov gap equations (which, quite consistently, shows that the
Gribov mass decreases with the increase of $r_{0}$). On the other hand, the
background gauge potentials considered in such works
would still allow a complete study of the semi-classical Gribov gap
equation (along the lines of the present analysis) but make extremely
difficult to construct explicit examples of Gribov copies. As we believe that,
when analyzing the Gribov problem with an external background field, it is
very instructive to analyze both the Gribov copies equation and, at the same
time, the corresponding Gribov gap equation (which, in a sense, are the two
sides of the same coin) we consider here the background gauge field in Eq.
\eqref{backgroundasMU}.

In this case, the LDW gauge fixing reads
\begin{equation}
\tilde{G}_{\mu}^{a}[B]=0\ ,
\end{equation}
so that the Gribov copies equation becomes
\begin{equation}
\partial^{\mu}A_{\mu}^{U}+g[B^{\mu},A_{\mu}^{U}]=0\ ,\label{masterequation}%
\end{equation}
where $A_{\mu}^{U}$ is defined in Eq. (\ref{backgroundsymmetry1}).
It is worth emphasizing that the background gauge field $B^{\mu}$ identically
satisfies the LDW gauge-fixing (as it should):%
\[
\partial^{\mu}B_{\mu}+g[B^{\mu},B_{\mu}]=0\ .
\]
The following standard parametrization of the $SU(2)$-valued functions
$U(x^{i})$ is useful
\begin{align}
U=Y^{0}\mathbf{1}+Y^{a}\tau_{a}  &  ,(Y^{0})^{2}+Y^{a}Y_{a}%
=1,\nonumber\label{functionsSU(2)}\\
(Y^{0})^{2}  &  +Y^{a}Y_{a}=1,
\end{align}
where $Y^{0}$ and $Y^{a}$ are functions on the coordinates $x^{i}$, and the
sum over repeated indices is understood also in the case of the group indices
(in which case the indices are raised and lowered with the flat metric
$\delta_{ab}$). The $SU(2)$ generators $\tau^{a}$ satisfy
\begin{equation}
\tau_{a}\tau_{b}=-\delta_{ab}\mathbf{1}-\epsilon_{abc}\tau_{c}%
\end{equation}
where $\mathbf{1}$ is the identity $2\times2$ matrix and $\epsilon_{abc}$ are the components of the totally antisymmetric Levi-Civita tensor with $\epsilon^{123}=\epsilon_{123}=1$.

\subsection{Gribov copies of the vacuum}

In the present case, the gauge transformations of the vacuum have the
expression (see Eq. (\ref{backgroundsymmetry1}))%
\[
0\rightarrow U^{-1}\partial_{\mu}U+\left(  U^{-1}B_{\mu}U-B_{\mu}\right)
\ .
\]
Correspondingly, the equation for the Gribov copies of the vacuum in the
presence of a background field reads
\begin{equation}
\partial^{\mu}\left(  U^{-1}\partial_{\mu}U+\left(  U^{-1}B_{\mu}U-B_{\mu
}\right)  \right)  +g[B^{\mu},U^{-1}\partial_{\mu}U+\left(  U^{-1}B_{\mu
}U-B_{\mu}\right)  ]=0\ . \label{vacuumcopies}%
\end{equation}
This, actually, is a system of coupled non-linear partial differential
equations. In order to reduce it consistently to a single differential
equation a particular hedgehog ansatz can be used \cite{nostro4} (see
Appendix \ref{chemical_gribov_copies} for the details on the vacuum case). This corresponds to the following ansatz for the gauge copy
\begin{equation}
U\ =\ Y^{0}(x^{i})\mathbf{1}+Y^{a}(x^{i})\tau_{a},\ , \label{expression-ansatz-hedgehog}%
\end{equation}
where
\begin{equation}
Y^{0}(x^{i})=\cos\alpha(x^{i}),\quad Y^{a}(x^{i})=\hat{n}^{a}%
\sin\alpha(x^{i}) \label{expression-y's}%
\end{equation}
being $\hat{n}_{a}$ normalized with respect to the internal metric
$\delta_{ab}$ as
\begin{equation}
\delta_{ab}\hat{n}^{a}\hat{n}^{b}=1\ .
\end{equation}%

\subsection{Vacuum Gribov copies with $T^{3}$ topology}

Let us analyze the Gribov copies equation in a flat spatial space with $T^{3}$-topology. Such choice of topology can be very useful in relation with
lattice studies \cite{luscher-paper,kogut}. We take the metric
\begin{equation}\label{T3_metric}
ds^{2}=\sum_{i=1}^{i=3}\lambda_{i}^{2}d^{2}\phi_{i},
\end{equation}
where the $\lambda_{i}\in\mathbb{R}$ represents the length of the torus along
the $i$-axis and the coordinates $\phi_{i}\in\lbrack0,2\pi)$ corresponds to
the $i-$th factor $S^{1}$ in $T^{3}$. In the $T^{3}$ case, the gauge transformation $U$ is independent of the Euclidean temporal coordinate $x^{0}$ and is \textit{proper} when \cite{nostro4}
\begin{equation}\label{strongboundary}
U(\phi_{i}+2m_{i}\pi)=U(\phi_{i}),\quad m_{i}\in\mathbb{Z},\quad
i=1,2,3.%
\end{equation}
The generalized hedgehog ansatz adapted to this topology reads
\begin{equation}\label{hedgehog_ansatz}
\alpha=\alpha(\phi_{1}),\quad\hat{n}^{1}=\cos(p\phi_{2}+q\phi_{3}),\quad
\hat{n}^{2}=\sin(p\phi_{2}+q\phi_{3}),\quad\hat{n}^{3}=0,
\end{equation}
with $p,q$ arbitrary integers. From this ansatz, the equation
\eqref{chemicalvacuumcopies} is reduce to the following single scalar
non-linear differential equation (see Appendix \ref{chemical_gribov_copies} for details),
\begin{equation}
\frac{d^{2}\alpha}{d\phi_{1}^{2}}=\xi\sin(2\alpha
),\label{chemicalsinglescalar}
\end{equation}
where
\begin{equation}\label{xi_definition}
\xi=\frac{\lambda_{1}^{2}}{2}\left( \frac{p^{2}}{\lambda_{2}^{2}}
+\frac{q^{2}}{\lambda_{3}^{2}}  +4\frac{r_{0}^{2}}{g}\right) ,
\end{equation}
and, according to \eqref{strongboundary}, the condition
\begin{equation}\label{alpha_condition}
\alpha(\phi_{1}+2\pi)=\alpha(\phi_{1})+2\pi k\;,\;k\in\mathbb{Z}
\end{equation}
must be fulfilled.
The equation \eqref{chemicalsinglescalar} can be reduced to a first order
conservation law
\begin{equation}\label{signxi}
V=\frac{1}{2}\left[  \left(  \frac{d\alpha}{d\phi_{1}}\right)  ^{2}+\xi
\cos(2\alpha)\right]  \Rightarrow\phi_{1}-\phi_{0}=\pm\int_{\alpha(\phi_{0}%
)}^{\alpha(\phi_{1})}\frac{dy}{\sqrt{2V-\xi\cos(2y)}} ,
\end{equation}
where $\phi_{0}$ and $V$ are integration constants. However, the integration
constant  $\phi_{0}$ is not relevant as it just corresponds to a shift of the
origin. Consequently, the relevant integration constants which labels different
solutions of the Gribov copies equation in Eq. (\ref{chemicalsinglescalar}) are $\xi$ and, through the boundary condition \eqref{strongboundary}, $V$ in Eq. (\ref{signxi}).

On the other hand, not any solution of Eq. (\ref{chemicalsinglescalar}) is an
allowed Gribov copy as the boundary conditions in Eq. (\ref{strongboundary})
must be required. Since $\phi_{1}$ belongs to the range $[0,2\pi)$, let us
take $\phi_{1}=2\pi$ and $\phi_{0}=0$. The condition \eqref{strongboundary}
implies $\alpha(\phi_{1})=\alpha(\phi_{0})+2\pi k$, where $k\in\mathbb{Z}$.
Taking this into account, we have for \eqref{signxi} the following expression
\begin{equation}
2\pi=\pm\frac{1}{\xi^{1/2}}\int_{\alpha(0)}^{\alpha(0)+2\pi k}\frac{dy}%
{\sqrt{Z-\cos(2y)}}\ ,\ \ Z=\frac{2V}{\xi}>1\ ,\label{fin1}%
\end{equation}
where $Z>1$ since the integrand must be well defined in the range $y\in(0,2\pi
k)$.

The present analysis shows that already the Gribov copies of the vacuum depend
very substantially on the background field. The different Gribov copies which
can be constructed with the present ansatz are in $1$-to-$1$ correspondence
with the solutions of Eq. \eqref{fin1}.

\begin{figure}
\begin{center}
\includegraphics[width=.57\textwidth]{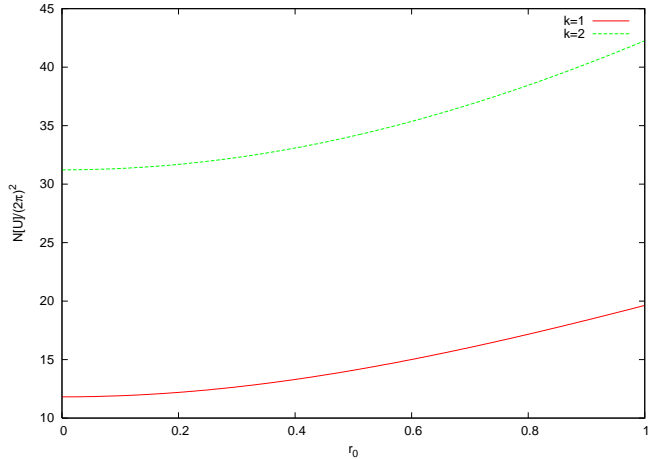}
\end{center}
\caption{The norm of the copies $\frac{N[U]}{(2\pi)^{2}}$, according to \eqref{norm_T3}, in the case $p=q=\lambda_{i}=1$ versus the background $r_{0}$ for $k=1$ (in red) and $k=2$ (in green). The solutions of $\alpha(\phi_{1})$ fulfil the condition \eqref{alpha_condition}.}%
\label{norm_copies}%
\end{figure}

As it is well known (see the detailed discussion
in \cite{normsorella}), the weight of a given copy $U$ is related to its norm
\begin{equation}\label{def_integral_norm}
N\left[U\right]= \int_{T^3} d^{4}x\sqrt{g}Tr\left[  \left(  U^{-1}\partial_{\mu}U+ U^{-1}B_{\mu} U -B_{\mu}\right)
^{2}\right],
\end{equation}
where in this case $g$ refers to the determinant of the metric associated to the line element \eqref{T3_metric}, setting the coupling constant to be zero.
In particular, the bigger is $N[U]$, the less relevant the copy is from the path integral point of view. As in this case there is a background potential, the integral \eqref{def_integral_norm} can be written as
\begin{eqnarray}\label{norm_T3}
N\left[U\right] &=& \frac{(2\pi)^{2}\lambda_{2}\lambda_{3}}{\lambda_{1}}\int_{0}^{2\pi} d\phi_{1}\left( \left(\frac{d\alpha}{d\phi_{1}}\right)^{2}+2\xi\sin^{2}\alpha\right)\nonumber \\
                &=&  \frac{(2\pi)^{2}\lambda_{2}\lambda_{3}}{\lambda_{1}}\int_{0}^{2\pi} d\phi_{1}\left(2V+3\xi\sin^{2}\alpha(\phi_{1})-\xi\cos^{2}\alpha(\phi_{1})\right) ,
\end{eqnarray}
where in the last equality we used the definition \eqref{signxi} of the constant $V$. In Figure \ref{norm_copies}, we show the norm $N[U]$ for $p=q=\lambda_{i}=1$ increases when $r_{0}$ grows both for $k=1$ and $k=2$, at least in the range $r_{0}\in(0.0,1.0)$, for solutions $\alpha(\phi_{1})$ such that fulfil the condition \eqref{alpha_condition} and $\alpha(0)=0$.
Consequently, in this region, the bigger $r_{0}$ the smaller the
importance of Gribov copies of the form considered here. It is necessary more computational power to see how is the behavior of the norm outside the region studied here (for instance, $|p|>1$ and $|q|>1|$). The above considerations suggest that the Gribov gap equation should also be
affected non-trivially by the background field. In the next section, it will
be shown that this is indeed the case.

\section{Solving the GZ Gap equation for $SU(2)$ with constant background field}

In order to determine the gap equation, we will proceed first to show the
effective potential to GZ action at one-loop approximation for the $SU(2)$
internal gauge group in the presence of a background potential discussed in Section \ref{GZ-background}. We will work at a small enough but non-zero temperature, taking into account the background field as
\begin{equation}\label{background_gap}
B_{\rho}^{a}=\frac{T}{g}r\delta^{a3}\delta_{\rho0},
\end{equation}
where $r$ is a dimensionless parameter related to the background field $r_{0}$ defined in \eqref{backgroundasMU} as $r=r_{0}/T$, with $T$ the temperature.
At a first glance, the above background field should have no physical at all as it is pure gauge. However, the gauge transformation which would remove it is not periodic in Euclidean time. Thus, such a gauge transformation is improper and it is not allowed. That is why it is conceptually important to include a non-vanishing temperature in the analysis of the gap equation. The present framework, despite its simplicity, it is still able to disclose in a very clean way the effects of the background gauge potential on the Gribov parameter. This approach is very close to the Polyakov-loop treatment \cite{polyakov-paper}, but in this case we will focus only on the dependence of the Gribov parameter with respect to the background parameter, keeping the temperature constant.

\begin{figure}
\begin{center}
\subfloat[][]{\includegraphics[width=.5\textwidth]{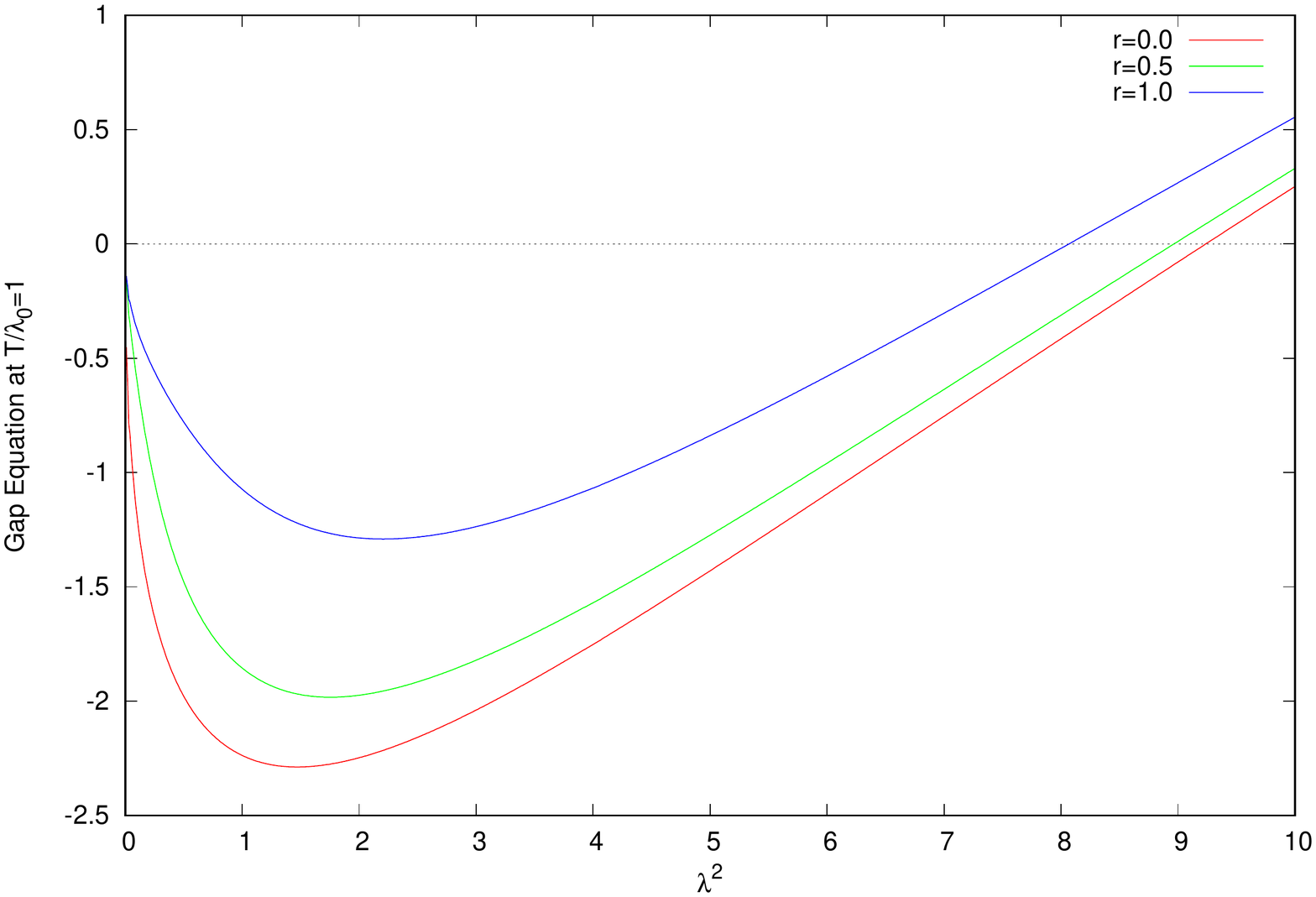}}
\subfloat[][]{\includegraphics[width=.5\textwidth]{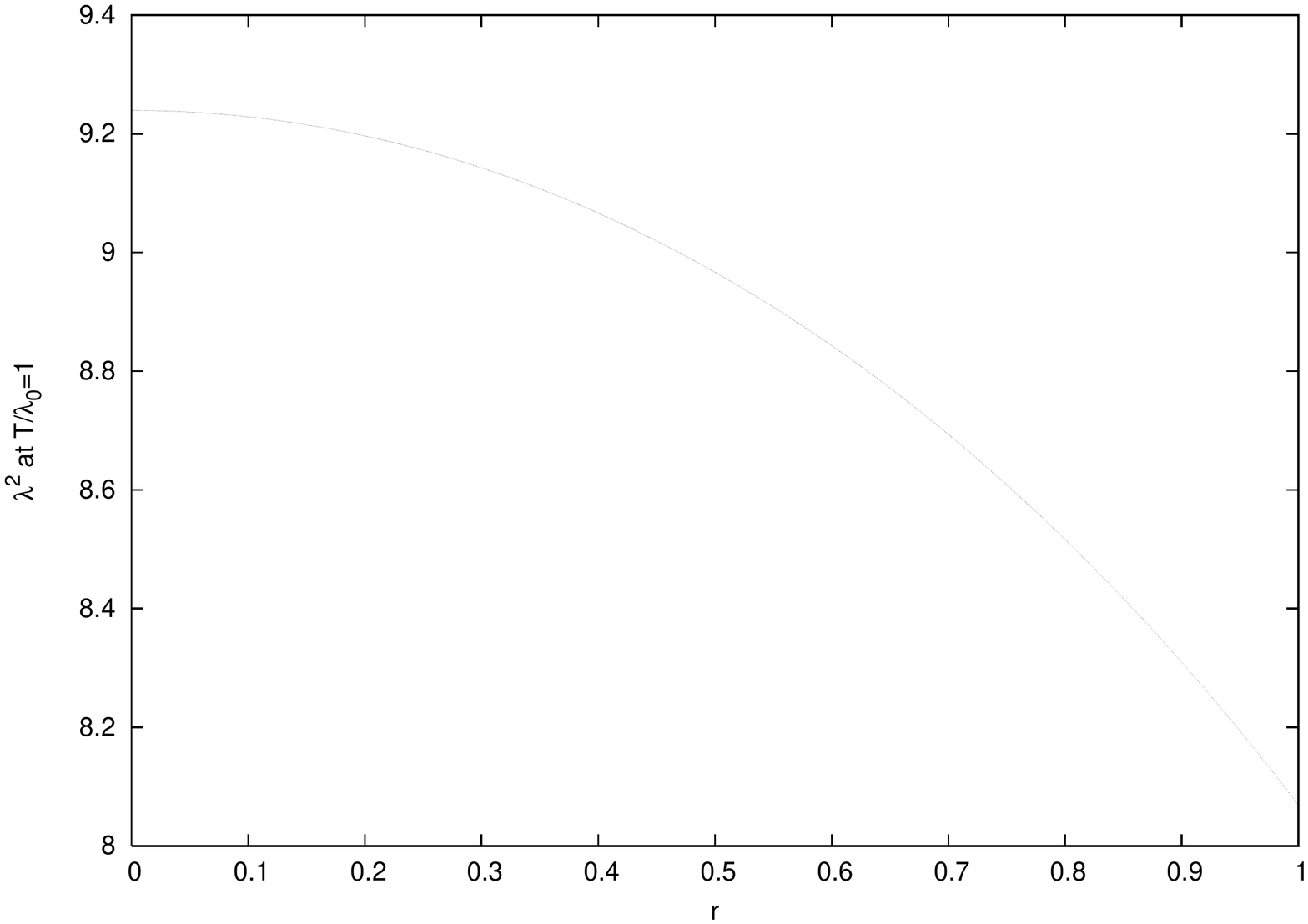}}
\end{center}
\caption{(a) The gap equation \eqref{gap_equationTnot0} as a function of $\lambda
^{2}$ for different values of $r$ at $T/\lambda_{0}=1$. The value of $\lambda$ which
corresponds the curve intersects the $y$- axis is the solution of
\eqref{gap_equationTnot0}. (b) The zeros of the gap equation \eqref{gap_equationTnot0} as a function of
the background field $r$ at $T/\lambda_{0}=1$. We see clearly the Gribov mass parameter decrease when $r$ in the range $[0,1]$.}
\label{fig_gap_equationTnot0}%
\end{figure}

In order to obtain the vacuum energy at one loop, we consider only from \eqref{backgroundactionGZ} the quadratic terms in the fields which are functionally integrated\footnote{Higher order corrections to \eqref{vacuum_potential} are obtained if we consider connected diagrams, see details in \cite{peskin}.}. We find \cite{polyakov-paper}
\begin{equation}
\varepsilon(r,\lambda^{2})=-d\frac{(N^{2}-1)}{2Ng^{2}}\lambda^{4}+(d-1)\frac{T}%
{2V}\text{Tr}\ln\frac{\overline{D}^{4}+\lambda^{4}}{\Lambda^{4}}-d\frac{T}%
{2V}\text{Tr}\ln\frac{-\overline{D}^{2}}{\Lambda^{2}},\label{vacuum_potential}%
\end{equation}
where $V$ is the Euclidean space volume, $\lambda^{4}=2Ng^{2}\gamma^{4}$, being
$\gamma$ is the Gribov parameter, $\overline{D}$ is the covariant background
derivative in the adjoint representation defined in \eqref{backgroundgaugefixingcondition}, and $\Lambda^{2}$ is a scale
parameter in order to regularize the result. We can rewrite
\eqref{vacuum_potential}, taking into account the Cartan subalgebra of $SU(2)$
is one-dimensional, as\footnote{We have taken the thermodynamic limit $V\to+\infty$ in equation \eqref{vacuum_potential_w_I}, implying that $\sum\limits_{q}\to V\int \frac{d^{3}q}{(2\pi)^{3}}$ \cite{notasdesorella}.}
\begin{equation}
\varepsilon(r,\lambda^{2})=-d\frac{(N^{2}-1)}{2Ng^{2}}\lambda^{4}+\frac{1}{2}(d-1)\sum\limits_{s=-1}^{s=1}\left[  I(sr,i\lambda^{2})+I(sr,-i\lambda^{2})\right]  -\frac{d}{2}\sum\limits_{s=-1}^{s=1}I(sr,0),\label{vacuum_potential_w_I}%
\end{equation}
where $s$ is the isospin $SU(2)$, and we defined the function
\begin{equation}
I(u,m^{2})=\frac{T}{V}\text{Tr}\ln\left(\frac{-\overline{D}^{2}+m^{2}}{\Lambda^{2}}\right)=T\sum\limits_{n=-\infty}^{+\infty}\int\frac{d^{3}q}{(2\pi)^{3}}\ln\left[\frac{(2\pi nT+uT)^{2}+\vec{q}^{2}+m^{2}}{\Lambda^{2}}\right]\;.\label{def_I_function}
\end{equation}
In the last definition, we expanded in the Fourier space the zero-component momentum in the Matsubara bosonic frequencies $2\pi nT$ \cite{lebellac,kapustabook}, and $\vec{q}$ denotes the spatial momentum vector. We will compute first
\eqref{def_I_function} using similar techniques which were already applied in
GZ approach \cite{polyakov-paper,thermo_compatibility}, which lead us the result (see details in Appendix \ref{detailed_computation_I})
\begin{equation}\label{I_function_result}
I(u,m^{2})=\frac{m^{4}}{32\pi^{2}}\left[  \ln\left(  \frac
{m^{2}}{\Lambda^{2}}\right)  -\frac{3}{2}\right]-\frac{T^{2}m^{2}}{\pi^{2}}\sum\limits_{n=1}^{+\infty}K_{2}\left(\frac{n\sqrt{m}}{T}\right)\frac{\cos(nu)}{n^{2}}\;.
\end{equation}
where $K_{\nu}$ is the modified Bessel function of the second kind extended to the complex plane \cite{ahlfors}.
Inserting \eqref{I_function_result} into \eqref{vacuum_potential_w_I}, we
have
\begin{eqnarray}\label{vacuum_potential_result}
\varepsilon(r,\lambda^{2})  & =&-\frac{d(N^{2}-1)\lambda^{4}}{2Ng^{2}}%
-3(d-1)\frac{\lambda^{4}}{32\pi^{2}}\left[  \ln\left(  \frac{\lambda^{2}}{\Lambda^{2}}\right)-\frac{3}{2}\right] \\
& & -\frac{i\lambda^{2}T^{2}}{2}(d-1)\sum\limits_{s=-1}^{s=1}\sum\limits_{n=1}^{+\infty}\left[K_{2}\left(\frac{n\sqrt{i\lambda^{2}}}{T}\right)-K_{2}\left(\frac{n\sqrt{-i\lambda^{2}}}{T}\right)\right]\frac{\cos(nrs)}{n^{2}}
-\frac{d}{2}\sum\limits_{s=-1}^{s=1}I(rs,0) \nonumber \;.
\end{eqnarray}
Because we are
interested in solving the gap equation \eqref{definition:gapequation}, we can neglect the last term of \eqref{vacuum_potential_result}, as it does not depend on $\lambda^{2}$.
In order to normalize the last equation, we shall choose $\Lambda^{2}$ such that for $T=0$ the solution is $\lambda_{0}=1$. Now, one can
write the gap equation \eqref{definition:gapequation} in the following way
\begin{equation}\label{gap_equationTnot0}
\frac{\partial\varepsilon_{T\neq0}(r,\lambda^{2})}{\partial\lambda^{2}}
-3(d-1)\frac{\lambda^{2}}{16\pi
^{2}}  \ln\left(  \frac{\lambda^{2}}{\lambda_{0}^{2}}\right)=0\;,
\end{equation}
where
\begin{equation*}
\varepsilon_{T\neq0}(r,\lambda^{2})=-\frac{i\lambda^{2}}{2}T^{2}(d-1)\sum\limits_{n=1}^{+\infty}\frac{\left[K_{2}\left(\frac{n\sqrt{i\lambda^{2}}}{T}\right)-K_{2}\left(\frac{n\sqrt{-i\lambda^{2}}}{T}\right)\right]}{n^{2}}\left[1+2\cos(nr)\right]\;.
\end{equation*}
The gap equation \eqref{gap_equationTnot0} can be solved using numerical techniques.
In Figure \ref{fig_gap_equationTnot0} (a), it is plotted the left hand side of the gap
equation \eqref{gap_equationTnot0} as a function of $\lambda^{2}$ for different
values of $r$.
We see the intersection value of the curve (which is the solution for a given value of $r$) decrease when the background $r$
grows, as it is shown more clearly in Figure \ref{fig_gap_equationTnot0} (b), where it
is shown the parameters $\lambda$ which are solution of gap equation at $T/\lambda_{0}=1$ versus $r$. We could interpret this as
the theory becomes less confined as the Gribov parameter reduces (see Section
\ref{conclusions}) at least in the range $r\in[0,1]$.
The present expression of the Gribov parameter is only valid as long as $\frac{\partial\lambda^{2}}{\partial r}\neq0$. The points where the latter derivative vanishes could signal a change on the phase diagram. Thus, when $\frac{\partial\lambda^{2}}{\partial\mu}\neq0$, the present semi-classical approximation is not valid anymore. Therefore, we have included the plots in Figure \ref{fig_gap_equationTnot0} only the region in which our approximation can be trusted.

\section{Conclusions and perspectives}
\label{conclusions}

In the present paper, it has been shown that the Gribov copies equation is
affected directly by the presence of a background gauge field. In particular,
explicit examples have been constructed in which the norm of the Gribov copies
satisfying the usual boundary conditions increases when the size of the
background field is very large.

The analysis of the semi-classical Gribov gap equation in the chosen
background gauge potential and of the dependence of the Gribov mass on the
background potential itself, quite consistently, confirms the above results.
Namely, we have shown that the larger is the size of the background gauge
potential, the smaller is the corresponding Gribov mass.

It is worth emphasizing the importance of the chosen constant background gauge
field is related to the fact that it appears in the analysis of the
computation of the Polyakov loop. Moreover, constant background gauge
potentials are very important also in relation with the non-Abelian Schwinger
effect \cite{schwinger2,schwinger3}. Although the constant gauge
potentials considered in that references allow a complete analysis of the
semi-classical Gribov gap equation, they make extremely difficult to construct
explicit examples of Gribov copies. We hope to come back on the more general
configurations considered in these works and on the
relations between the non-Abelian Schwinger effect and the Gribov problem in
background gauge fields in a future publication.

\subsection*{Acknowledgements}

The authors would like to thank Patricio Salgado-Rebolledo for very useful
conversations and remarks on the topics of this work. We wish also thank David Dudal and David Vercauteren for constructive criticism and important suggestions on delicate issues analyzed in this paper. This work is supported
by Fondecyt Grant 1160137. P.P. was partially supported by Fondecyt Grant
1140155 and also thanks the Faculty of Mathematics and Physics
of Charles University in Prague, Czech Republic, for the kind hospitality
during part of the development of this work. The Centro de Estudios Cient\'{\i}ficos (CECs) is funded by the
Chilean Government through the Centers of Excellence Base Financing Program of Conicyt.

\appendix

\section{Gribov copies with a constant background field}
\label{chemical_gribov_copies}

In this appendix we consider the derivations
and properties of the equation of gauge-equivalent fields satisfying the LDW
gauge in the presence of background field.
Our aim is to calculate the condition for existence of Gribov copies in the
vacuum. Thus, we must compute the following expression
\begin{align*}
U^{-1}\partial_{\mu}U+U^{-1}B_{\mu}U-B_{\mu} &  =\left(Y^{0}\partial_{\mu}Y^{c}-Y^{c}\partial_{\mu}Y^{0}+\epsilon_{abc}Y_{a}\partial_{\mu}Y_{b}\right)\tau_{c}\nonumber\\
&  -\frac{2r_{0}}{g}\delta_{\mu0}\left(\epsilon_{a3c}Y^{a}Y^{0}+Y^{3}
Y^{c}\right)\tau_{c}+\frac{2r_{0}}{g}\delta_{\mu0}Y^{a}Y_{a}\tau_{3}.
\end{align*}
The next step is to apply to this last expression the covariant background
derivative and set it to be zero according to \eqref{masterequation}. This results in the following expression {\small
\begin{eqnarray}\label{chemicalvacuumcopies}
& &  \left( -Y^{0}\square Y^{c} - Y^{c}\square Y^{0} \epsilon_{abc}Y^{a}\square Y^{b} \right)\tau_{c}-\frac{2r_{0}}{g}\left(Y^{c}\dot{Y}^{3}+\dot{Y}^{c}Y^{3}+\epsilon_{a3c}\left[Y^{0}\dot{Y}^{a}+\dot{Y}^{0}Y^{a}\right]\right)\tau_{c}  \nonumber\\
& &\left[-2r_{0}\epsilon_{3bc}\left(Y^{0}\dot{Y}^{b}-\dot{Y}^{0}Y^{b}-\dot{Y}^{c}Y^{3}+Y^{c}\dot{Y}^{3}\right)  -4\frac{r_{0}^{2}}{g}\left(Y^{0}Y^{c}+\epsilon_{3bc}Y^{3}Y^{b}\right)\right]\tau_{c}+\frac{4r_{0}}{g}\dot{Y}^{a}Y_{a}\tau_{3}=0 ,
\end{eqnarray}}
where $\square(\ldots)=\partial_{\mu}\partial^{\mu}(\ldots)$, and de dot represents the derivative with respect to the component which the background field belongs.
In the particular case of flat spatial space $T^{3}$ for the $Y^{\mu}$ prescription \eqref{expression-y's}, for the hedgehog ansatz \eqref{hedgehog_ansatz} the set of equations \eqref{chemicalvacuumcopies} reduces to these three equations
\begin{equation*}
\hat{n}^{c}\square\alpha+\frac{1}{2}\sin(2\alpha)\square\hat{n}^{c}-\frac{2r_{0}^{2}}{g}\sin(2\alpha)\hat{n}^{c}=0.
\end{equation*}
Taking into account the $T^{3}$-metric \eqref{T3_metric}, we end up with the following ordinary differential equation
\begin{equation*}
\frac{d^{2}\alpha}{d\phi_{1}^{2}}-\frac{\beta(p,q)}{2}\sin(2\alpha)-\frac{2r_{0}^{2}\lambda_{1}^{2}}{g}\sin(2\alpha)=0,
\end{equation*}
where we defined $\beta(p,q)=\lambda_{1}^{2}\left(\frac{p^{2}}{\lambda_{2}^{2}}+\frac{q^{2}}{\lambda_{3}^{2}}\right)$. If we introduce the $\xi$ definition \eqref{xi_definition}, we get the result \eqref{chemicalsinglescalar}.

\section{Computation of the I-function}

\label{detailed_computation_I}

In this Appendix we will derive in some detail the equation
\eqref{I_function_result}, following the lines of \cite{polyakov-paper,thermo_compatibility}. The quantity we would like to compute is
\begin{equation*}
I(u,m^{2})=T\sum\limits_{n=-\infty}^{+\infty}\int\frac{d^{3}q}{(2\pi)^{3}}\ln\left[\frac{(2\pi nT+uT)^{2}+\vec{q}^{2}+m^{2}}{\Lambda^{2}}\right] ,
\end{equation*}
where $u$ is related to the background field, $m^{2}$ is a squared mass (that could be
complex), and $\Lambda^{2}$ is a quantity we used to regularize the divergence. We can write $I(u,m^{2})$ as the derivative respect of some auxiliary variable
$\epsilon$, and then taking the limit $\epsilon\rightarrow0$:
\begin{equation*}
I(u,m^{2})=-\lim_{\epsilon\rightarrow0}\frac{\partial}{\partial\epsilon
}\left(T\sum\limits_{n=-\infty}^{+\infty}\int\frac{d^{3}q}{(2\pi)^{3}}\left(\frac{(2\pi nT+uT)^{2}+\vec{q}^{2}+m^{2}}{\Lambda^{2}}\right)  ^{-\epsilon}\right)\;.
\end{equation*}
Defining a new variable $t$ as $|\vec{q}|=t\sqrt{(2\pi nT+uT)^{2}+m^{2}}$ and
passing to spherical coordinates, we have
\begin{equation*}
I(u,m^{2})=-\lim_{\epsilon\rightarrow0}\frac{\partial}{\partial\epsilon
}\left(T\sum\limits_{n=-\infty}^{+\infty}  \frac{\Lambda^{2\epsilon}}{2\pi^{2}}\left((2\pi nT+uT)^{2}
+m^{2}\right)  ^{3/2-\epsilon}\int_{0}^{+\infty}dtt^{2}(1+t^{2})^{-\epsilon
}\right)  \;.
\end{equation*}
We can write the last integral as
\begin{equation*}
\int_{0}^{+\infty}dtt^{2}(1+t^{2})^{-\epsilon}=\frac{\sqrt{\pi}}{4}\frac{\Gamma(\epsilon-3/2)}{\Gamma(\epsilon)}\;,
\end{equation*}
where in the equality we take the analytical
continuation because strictly must be $Re(\epsilon)>3/2$ in the real domain. So,
using the definition of Gamma function and making another change of the integration variable, we have
\begin{equation*}
I(u,m^{2})=-\lim_{\epsilon\rightarrow0}\frac{\partial}{\partial\epsilon
}\left(  \frac{T^{4-2\epsilon}\Lambda^{2\epsilon}}{2^{2\epsilon}\pi^{2\epsilon-3/2}\Gamma(\epsilon)}\sum\limits_{n=-\infty}^{+\infty}\int_{0}^{+\infty}dyy^{\epsilon-5/2}e^{-y\left[v^{2}+(n+c)^{2}\right]}\right)\;,
\end{equation*}
where we defined $v^{2}=\frac{m^{2}}{4\pi^{2}T^{2}}$ and $c=\frac{u}{2\pi}$.
Remembering the Poisson summation formula (valid for positive $y$), we can rewrite the sum over $n$ as
\begin{equation*}
\sum\limits_{n=-\infty}^{+\infty}e^{-y(n+c)^{2}}=\sqrt{\frac{\pi}{y}}\left(1+2\sum\limits_{n=1}^{+\infty}e^{\frac{-n^{2}\pi^{2}}{y}}\cos(2\pi nc)\right)\;.
\end{equation*}
We shall compute first the $n=0$ mode $I_{n=0}(u,m^{2})$, which can be done using the Gamma function properties, making a change of variable $z=v^{2}y$ and performing the limit $\epsilon\to0$,
\begin{equation}\label{I_n0}
I_{n=0}(u,m^{2})=-\lim_{\epsilon\rightarrow0}\frac{\partial}{\partial\epsilon
}\left(  \frac{T^{4-2\epsilon}\Lambda^{2\epsilon}(v^{2})^{2-\epsilon}}{2^{2\epsilon}\pi^{2\epsilon-2}\Gamma(\epsilon)}\int_{0}^{+\infty}dzz^{\epsilon-3}e^{-z}\right)=\frac{m^{4}}{32\pi^{2}}\left[\ln\left(\frac{m^{2}}{\Lambda^{2}}\right)-\frac{3}{2}\right]\;,
\end{equation}
which coincides with the usual result at $T=0$ \cite{peskin}.

The $n\neq0$ modes $I_{n\neq0}(u,m^{2})$ are
\begin{equation*}
I_{n\neq0}(u,m^{2})=-\lim_{\epsilon\rightarrow0}\frac{\partial}{\partial\epsilon
}\left(  \frac{T^{4-2\epsilon}\Lambda^{2\epsilon}(v^{2})^{2-\epsilon}}{2^{2\epsilon}\pi^{2\epsilon-2}\Gamma(\epsilon)}\sum\limits_{n=1}^{+\infty}\cos(nu)\int_{0}^{+\infty}dzz^{\epsilon-3}e^{-z-\frac{n^{2}\pi^{2}v^{2}}{z}}\right)\;,
\end{equation*}
The last integral can be solved in terms of the modified Bessel function of the second kind, taking into account the property \cite{bessel_second}
\begin{equation*}
\int_{0}^{+\infty}dxx^{-\nu-1}e^{-x-\frac{b}{x}}=\frac{2}{b^{\nu/2}}K_{\nu}(2\sqrt{b}),
\end{equation*}
leading to
\begin{equation}\label{I_not0}
I_{n\neq0}(u,m^{2})=-\lim_{\epsilon\rightarrow0}\frac{\partial}{\partial\epsilon
}\left(  \frac{T^{4-2\epsilon}\Lambda^{2\epsilon}(v^{2})^{\frac{2-\epsilon}{2}}}{2^{2\epsilon-2}\pi^{2\epsilon}\Gamma(\epsilon)}\sum\limits_{n=1}^{n=+\infty}\frac{\cos(nu)}{n^{2-\epsilon}}K_{2-\epsilon}\left(\frac{n\sqrt{m^{2}}}{T}\right)\right)=-\frac{T^{2}m^{2}}{\pi^{2}}\sum\limits_{n=1}^{n=+\infty}K_{2}\left(\frac{n\sqrt{m^{2}}}{T}\right)\frac{\cos(nu)}{n^{2}}\;,
\end{equation}
where we took the limit in the last equality using the property $\Gamma(\epsilon)=\frac{1}{\epsilon}-\gamma-\mathcal{O}(\epsilon)$ when $\epsilon\to0$, being $\gamma$ the  Euler-Mascheroni constant.
Combining the results \eqref{I_n0} and \eqref{I_not0}, we arrive to the desired formula \eqref{I_function_result}.

\end{document}